\documentclass[trackchanges]{aastex701}

\begin{document}
\title{Local Intensity-based Correction for SMFT Longitudinal Magnetograms:
  Validation with SDO/HMI Observations}

\author[0000-0002-1396-7603]{Liu, S.}
\affiliation{State Key Laboratory of Solar Activity and Space Weather,
  National Astronomical Observatories, Chinese Academy of Sciences,
  Beijing 100101, China}
\affiliation{School of Astronomy and Space Sciences,
  University of Chinese Academy of Sciences, Beijing 100049, China}
\email{sjt@nao.cas.cn}

\author[0000-0003-4244-1077]{Xu, H.Q.}
\affiliation{State Key Laboratory of Solar Activity and Space Weather,
  National Astronomical Observatories, Chinese Academy of Sciences,
  Beijing 100101, China}
\affiliation{School of Astronomy and Space Sciences,
  University of Chinese Academy of Sciences, Beijing 100049, China}
\email{}

\author[0000-0002-5152-7318]{Su, J.T.}
\affiliation{State Key Laboratory of Solar Activity and Space Weather,
  National Astronomical Observatories, Chinese Academy of Sciences,
  Beijing 100101, China}
\affiliation{School of Astronomy and Space Sciences,
  University of Chinese Academy of Sciences, Beijing 100049, China}
\email{}

\begin{abstract}
The Solar Magnetic Field Telescope (SMFT) at Huairou Solar Observing Station (HSOS) derives photospheric magnetic fields from single-wavelength Stokes measurements using a linear calibration based on the weak-field approximation. In sunspot umbrae, this linear approach systematically underestimates the true magnetic field strength---a well-known effect of magnetic underestimation. To mitigate this bias, we apply the Local Intensity-based Correction (LIC) as implemented in the standard HSOS data reduction procedure and validate its performance against SDO/HMI full-disk longitudinal magnetograms that are co-temporal and co-spatial with the SMFT observations for 26 active regions observed in 2019 and 2024. Pixel-level alignment between SMFT images ($992\times830$, $0.242^{\prime\prime}$~pixel$^{-1}$) and HMI snapshots ($4096\times4096$, $0.505^{\prime\prime}$~pixel$^{-1}$) is achieved using SIFT feature matching and RANSAC-based homography estimation, yielding inlier fractions of 40\% to 88\%. The LIC correction consistently improves the Pearson correlation coefficient $r$ with respect to HMI for all these 26 regions, with a mean improvement of $\Delta r = +0.041$ and a maximum of $\Delta r = +0.197$ for AR24114 (NOAA 13667). Residual discrepancies persist in the most severely underestimated pixels, highlighting an inherent limitation of any single-wavelength linear calibration. Nevertheless, our results demonstrate that the LIC correction is a robust and effective preprocessing step, and we recommend its routine application in quantitative analyses of SMFT longitudinal magnetograms.
\end{abstract}



\section{Introduction}
\label{sec:intro}

The Solar Magnetic Field Telescope (SMFT) at Huairou Solar Observing Station
(HSOS) has provided routine vector magnetic field measurements of the Sun since
the 1980s \citep{Ai1986}.
The instrument performs narrow-band filtergram observations at a single
wavelength position near the Fe\,{\sc i}\,5324.19\,\AA\ photospheric spectral
line (Land\'e factor $g=1.5$, equivalent width $\approx0.33$\,\AA).
From the four Stokes parameters $I$, $Q$, $U$, and $V$, we derive the longitudinal
and transverse field strengths via linear calibration under the weak-field
approximation following \citet{1989ApJ...343..920J,1991ApJ...372..694J}:
\begin{equation}
B_L = C_L\,V, \label{eq:Bl}
\end{equation}
\begin{equation}
B_T = C_T\,\bigl(Q^2+U^2\bigr)^{1/4}, \label{eq:Bt}
\end{equation}
where the field inclination and azimuth are further obtained as
\begin{equation}
\theta = \arctan\bigl(B_T/B_L\bigr), \label{eq:theta}
\end{equation}
\begin{equation}
\phi = \frac{1}{2}\arctan\bigl(U/Q\bigr). \label{eq:phi}
\end{equation}
The calibration normal used constants are $C_L$=8381 G and $C_T$=6790 G \citep{Ai1986,1996ApJ...456..861W,2004ChJAA...4..365S}.
Because the measurement is performed at a single wavelength point rather than
across the full spectral profile, the Stokes $V$ signal  becomes progressively
nonlinear with respect to $B_{\parallel}$ in strong-field regions; the linear
calibration therefore systematically underestimates the true field strength,
an effect known as magnetic underestimation.

This underestimation effect has been extensively documented. \citet{2007PASJ...59S.619C} successfully tested the $I$--underestimation anti-correlation on Hinode/SP and /NFI data, demonstrating that it can be used to calibrate line-of-sight magnetograms even in dark umbrae. \citet{2007PASJ...59S.625M} confirmed the same anti-correlation (Zeeman saturation effect) through cross-calibration of Hinode SP vector magnetograms with MDI data.

Various approaches have been developed to address this nonlinear underestimation in strong-field regions. \citet{2004ChJAA...4..365S} developed multi-wavelength procedures that mitigate the problem by scanning multiple wavelength positions across the Fe~I 5324.18~\AA\ line, but these require more wavelength points than the routine single-point SMFT mode. For multi-wavelength spectropolarimetric data from other instruments, \citet{2014MNRAS.445...49B} performed full Stokes-profile inversion using a Milne--Eddington atmosphere model with six-point spectral sampling, which effectively recovers the true field but requires full spectral profile information not available in routine SMFT observations. More relevant to the HSOS context, \citet{2021SoPh..296..165P} established a global nonlinear $V/I$ vs $B_z$ relationship by cross-calibrating SMFT data with SDO/HMI, providing a practical correction method that does not require multi-wavelength scanning. However, this approach applies a single functional form to all active regions, implicitly assuming a universal relationship across different solar conditions and relying on external HMI data. Despite their differences, all these studies demonstrate that the underestimation bias becomes severe in the darkest umbrae.
\citet{2021RAA....21...67X} adopted the $I$--underestimation anti-correlation and used a second-order
polynomial, achieving an 88\% correction rate, while it requires careful treatment of the
boundary region where the threshold is strictly defined.
To overcome these limitations in HSOS data processing, we propose a local intensity-
based correction (LIC) that uses $I$ itself as a pixel-wise proxy:

\begin{equation}
\mathrm{CF} = \frac{\langle I\rangle}{I}, \label{eq:cf}
\end{equation}
and the corrected longitudinal field is
\begin{equation}
B_L = B_{L0}\times\mathrm{CF}, \label{eq:bl_corr}
\end{equation}
where $B_{L0}$ is the linearly calibrated field from the routine
procedure and $\langle I\rangle$ is the spatial mean of the continuum intensity $I$ over the full field of view of the SMFT image. This normalization by the global mean intensity is analogous in spirit to the $I_{\rm max}$ normalization used in previous studies \citep{2021RAA....21...67X,2021SoPh..296..165P}, but differs fundamentally in that the correction is applied pixel-wise rather than through a globally defined function. Unlike the approach of \citet{2021SoPh..296..165P}, which fits a single global nonlinear $V/I$ vs $B_z$ relationship for all active regions using HMI cross-calibration, LIC adapts to the local conditions of each umbra, penumbra, or quiet-Sun region independently, requiring no external reference, no global fitting, and no explicit threshold. This pixel-wise formulation makes it particularly suitable for automated processing of historical HSOS data for which contemporaneous HMI or other reference magnetograms are not available.

In this paper we validate the LIC against co-temporal SDO/HMI
full-disk longitudinal magnetograms for 26 active regions observed in 2019 (near solar minimum) and 2024 (near solar maximum).
We first establish pixel-level spatial registration between SMFT ($992\times830$,
$0.242^{\prime\prime}$~pixel) and HMI snapshots ($4096\times4096$, $0.505^{\prime\prime}$~pixel) using
Scale-Invariant Feature Transform (SIFT) keypoint matching followed by Random
Sample Consensus (RANSAC) homography estimation.  After alignment, we compare $B_{L0}$, $B_L$, and the HMI reference for all 26 regions to quantify the improvement in correlation achieved by the LIC correction.

The paper is organized as follows.
Section~\ref{sec:data} describes the SMFT and HMI data, the LIC
procedure, and the image registration method.
Section~\ref{sec:results} presents the statistical validation and qualitative
comparisons for representative active regions.
Section~\ref{sec:discuss} discusses the implications and limitations.
Section~\ref{sec:conclude} summarizes our main findings.

\section{Data and Method}
\label{sec:data}

\subsection{SMFT observations}

The SMFT at HSOS uses a 35\,cm aperture vector magnetograph with a
birefringent filter tuned to the Fe\,{\sc i}\,5324.19\,\AA\ photospheric
spectral line (Land\'e factor $g=1.5$, equivalent width $\approx0.33$\,\AA).
The instrument performs narrow-band filtergram observations at two
wavelength positions: the longitudinal field is derived from Stokes $V$
measurements taken at 0.075\,\AA\ in the blue wing of the line center,
while the transverse field is derived from Stokes $Q$ and $U$ measurements
taken at the line center.  The observations are made with a KD*P electro-optic
crystal modulator that switches between left- and right-circular polarization
states \citep{Ai1986}.  For each active region, we perform
polarization demodulation to obtain the Stokes parameters:
\begin{equation}
Q = \frac{I_{Q+} - I_{Q-}}{I_{Q+} + I_{Q-}}, \quad
U = \frac{I_{U+} - I_{U-}}{I_{U+} + I_{U-}}, \quad
V = \frac{I_{V+} - I_{V-}}{I_{V+} + I_{V-}},
\end{equation}
where $I_{V+}$ and $I_{V-}$ are the intensities measured in the two
circular polarization states, and similarly for $Q$ and $U$.

The continuum intensity $I$ is obtained from the
sum of the left- and right-circular polarization intensities:
\begin{equation}
I = I_{V+} + I_{V-}.
\end{equation}
We use standard SMFT data stored as multi-layer FITS
files ($992\times830$ pixels, $0.242^{\prime\prime}$~pixel$^{-1}$, with a field of view
of approximately $240^{\prime\prime}\times200^{\prime\prime}$).  The linearly calibrated longitudinal field is obtained via
Equation~(\ref{eq:Bl}), with the calibration constant $C_L = 8381$\,G
\citep{1996ApJ...456..861W,2004ChJAA...4..365S}.  The correction procedure
that produces $B_L$ from $B_{L0}$ is described in detail in
Section~\ref{sec:lic}.

A total of 26 active-region observations acquired in 2019 and 2024 are used,
spanning 25 distinct active regions (AR24124 was observed on two separate
days).  The sample ranges from weak simple configurations to complex
multi-spot groups as listed in Table~\ref{tab1}.

\begin{table}
\centering
\caption{List of 26 active-region observations used in this study}
\label{tab1}
\scriptsize
\begin{tabular}{lccccc}
\hline
HSOS AR & NOAA AR & SMFT obs (TAI) & HMI obs (TAI) & Lat ($^\circ$) & Lon ($^\circ$) \\
\hline
AR19007 & NOAA12738 & 2019-04-14 03:20:27 & 2019-04-14 03:24:00 & +6 & -19.00 \\
AR19010 & NOAA12741 & 2019-05-14 02:26:52 & 2019-05-14 02:24:00 & +5 & -29.00 \\
AR24040 & NOAA13583 & 2024-02-13 02:41:44 & 2024-02-13 02:36:00 & +9  & -16.11 \\
AR24047 & NOAA13590 & 2024-02-23 02:10:22 & 2024-02-23 02:12:00 & +19 & -16.72 \\
AR24051 & NOAA13595 & 2024-03-01 06:39:34 & 2024-03-01 06:36:00 & +20 & +11.95 \\
AR24069 & NOAA13614 & 2024-03-23 04:05:25 & 2024-03-23 04:00:00 & +18 &  -6.54 \\
AR24108 & NOAA13661 & 2024-05-03 09:21:26 & 2024-05-03 09:24:00 & +22 & -36.37 \\
AR24114 & NOAA13667 & 2024-05-09 04:33:36 & 2024-05-09 04:36:00 & +28 & -18.69 \\
AR24122 & NOAA13674 & 2024-05-16 07:31:07 & 2024-05-16 07:36:00 & -14 &  -7.49 \\
AR24124 & NOAA13679 & 2024-05-15 07:25:20 & 2024-05-15 07:24:00 &  -9 & -43.77 \\
AR24124 & NOAA13679 & 2024-05-17 02:26:36 & 2024-05-17 02:24:00 &  -9 & -20.05 \\
AR24128 & NOAA13685 & 2024-05-20 06:36:59 & 2024-05-20 06:36:00 & -13 & -28.08 \\
AR24145 & NOAA13702 & 2024-06-10 01:03:35 & 2024-06-10 01:00:00 & +18 & +18.75 \\
AR24153 & NOAA13712 & 2024-06-15 06:35:42 & 2024-06-15 06:36:00 & -15 & -31.02 \\
AR24160 & NOAA13719 & 2024-06-24 02:02:11 & 2024-06-24 02:00:00 & -13 & -14.40 \\
AR24178 & NOAA13738 & 2024-07-11 04:05:06 & 2024-07-11 04:00:00 &  -9 &  -2.26 \\
AR24193 & NOAA13756 & 2024-07-22 03:20:09 & 2024-07-22 03:24:00 & -18 & +11.88 \\
AR24199 & NOAA13763 & 2024-07-28 00:17:37 & 2024-07-28 00:12:00 &  +3 &  +5.66 \\
AR24216 & NOAA13781 & 2024-08-10 01:47:00 & 2024-08-10 01:48:00 & +14 & -23.69 \\
AR24221 & NOAA13784 & 2024-08-14 07:29:04 & 2024-08-14 07:24:00 & +15 &  -0.67 \\
AR24227 & NOAA13790 & 2024-08-22 07:15:32 & 2024-08-22 07:12:00 & -14 &  +9.94 \\
AR24230 & NOAA13794 & 2024-08-22 03:05:03 & 2024-08-22 03:00:00 & +18 & +22.65 \\
AR24235 & NOAA13799 & 2024-08-27 02:24:11 & 2024-08-27 02:24:00 & -10 &  +5.33 \\
AR24251 & NOAA13814 & 2024-09-12 04:47:49 & 2024-09-12 04:48:00 & +15 & +16.98 \\
AR24274 & NOAA13848 & 2024-10-08 06:07:26 & 2024-10-08 06:12:00 & +15 &  +0.86 \\
AR24299 & NOAA13879 & 2024-11-03 02:39:24 & 2024-11-03 02:36:00 & +18 & -20.13 \\
\hline
\end{tabular}
\end{table}

\noindent {\footnotesize The HSOS AR designation follows the internal naming convention of Huairou Solar Observing Station. AR24124 (NOAA 13679) was observed on two separate days (May 15 and May 17). HMI obs denotes the time of the HMI 720-second cadence magnetogram closest to the SMFT observation, with the time difference typically within $\pm 1$ minute. The heliographic coordinates (Lat and Lon) are taken from the HSOS observation.}

\subsection{Local Intensity-based Correction}
\label{sec:lic}

The Local Intensity-based Correction (LIC) is applied pixel-wise following
the definition in Equation~(\ref{eq:cf}), where the correction factor
$\mathrm{CF} = \langle I\rangle / I$ is inversely proportional to the
local continuum intensity.  The corrected longitudinal field is then
obtained as $B_L = B_{L0} \times \mathrm{CF}$ [Equation~(\ref{eq:bl_corr})].
This correction factor is greater than unity in dark umbrae ($I < \langle I\rangle$),
recovering the underestimated field, and less than unity in bright features
($I > \langle I\rangle$), preventing overcorrection.

The LIC is implemented in the HSOS standard procedure \texttt{lic.pro}.
Its physical basis is the empirical anti-correlation between continuum
intensity and the apparent $V/I$ underestimation level: in dark sunspot
umbrae $I$ is well below the quiet-Sun average while the $V$ signal is
simultaneously reduced by the onset of underestimation.  The method requires
no multi-wavelength scanning, no reference magnetogram, and no explicit
underestimation threshold, making it suited for routine automated processing.

\subsection{SDO/HMI data}

The Helioseismic and Magnetic Imager (HMI) aboard the Solar Dynamics Observatory (SDO) provides full-disk longitudinal magnetograms at a nominal cadence of 720\,s \citep{2012SoPh..275..229S}.  Each magnetogram has $4096\times4096$ pixels at a plate scale of $0.505^{\prime\prime}$~pixel$^{-1}$, corresponding to a spatial resolution of approximately $1^{\prime\prime}$ on the solar surface, with a typical noise level of approximately 6.3\,G per pixel \citep{2012SoPh..279..295L}.  The HMI instrument observes the Fe\,{\sc i}\,6173\,\AA\ photospheric spectral line using six wavelength positions across the line profile, acquiring filtergrams in multiple polarization states.  The line-of-sight magnetic field is derived from the circular polarization measurements via a well-established calibration pipeline \citep{2012SoPh..275..229S}.  Unlike the SMFT single-wavelength linear calibration, which systematically underestimates strong fields, the HMI longitudinal magnetograms are based on multi-wavelength spectral information and are therefore suitable as an independent reference for validation.

The HMI data used in this work belong to the series \texttt{hmi.M\_720s} and are downloaded from the Joint Science Operations Center (JSOC) at Stanford University via the data export interface.  For each SMFT active-region observation, we select the HMI full-disk magnetogram closest in time (typically within $\pm 1$\,minute as listed in Table~\ref{tab1}) and extract a subframe corresponding to the SMFT field of view after image registration.  The full-disk data are stored in FITS format with complete header metadata providing essential observational parameters.

\subsection{SMFT--HMI image registration}

To enable pixel-wise quantitative comparison between SMFT active region
($992\times830$, $0.242^{\prime\prime}$/pixel) and HMI full-disk magnetograms
($4096\times4096$, $0.505^{\prime\prime}$/pixel), we perform feature-based image
registration in two stages.
Stage~1 removes the off-disk background from the HMI magnetogram by cropping
to a solar disc of radius $R_{\odot}=1600$ pixels.
Stage~2 detects and matches SIFT keypoints \citep{Lowe2004} between
the SMFT cutout and the corresponding HMI subframe.
The SIFT detector is configured with
${\rm contrastThreshold}=0.02$, ${\rm edgeThreshold}=5$, ${\rm sigma}=1.0$,
and ${\rm nFeatures}=50\,000$ to maximize the number of keypoints in the
low-contrast SMFT magnetograms.
Matches are filtered with Lowe's ratio test ($k=2$, threshold 0.75).
An affine homography is estimated with RANSAC
(${\rm reprojectionThreshold}=10$ pixels, $N=5000$ iterations).
The SIFT algorithm is inherently invariant to image translation, rotation,
and scaling, making it well suited for registering SMFT and HMI magnetograms
that may have different orientations, plate scales, and field-of-view coverage.
The affine transform is applied to the HMI subframe using bilinear
interpolation and inverse mapping to produce a registered image that exactly
matches the SMFT dimensions.

\section{Results}
\label{sec:results}

\begin{figure}
\centering
\includegraphics[width=\linewidth]{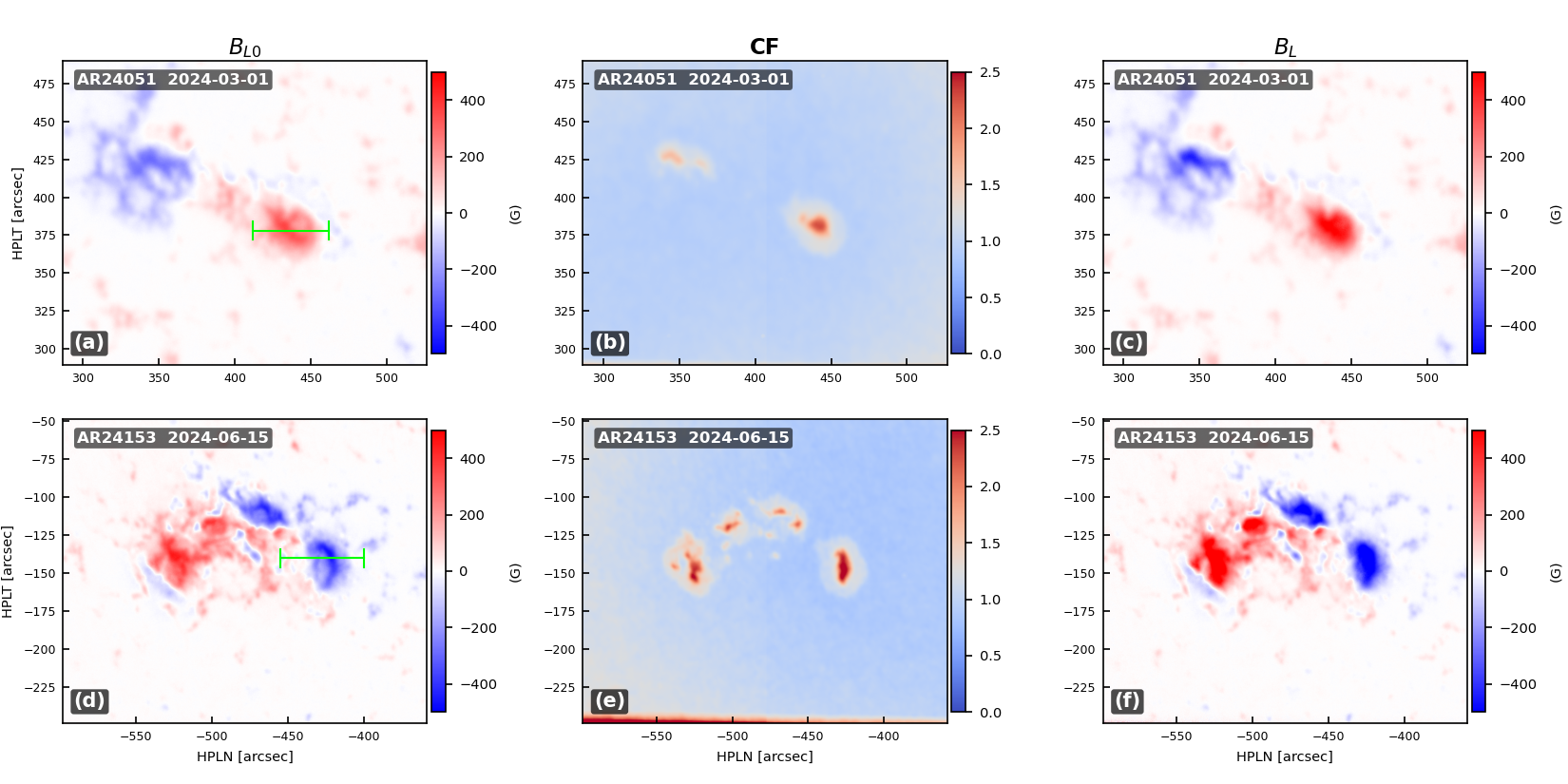}
\caption{Comparison of the uncorrected field $B_{L0}$ (left), the correction
factor $CF$ (middle), and the corrected field $B_L$ (right) for AR24051 (top)
and AR24153 (bottom). The lime segments in the left column mark the
horizontal cuts used in Figure~\ref{fig:linecut}.}
\label{fig:bl_cf}
\end{figure}

We first examine the effect of the LIC correction on individual active regions. Figure~\ref{fig:bl_cf} compares the uncorrected field $B_{L0}$ (left), the correction factor $CF$ (middle), and the corrected field $B_L$ (right) for AR24051 (top, a relatively weak active region) and AR24153 (bottom, a strong active region). In the dark umbrae where the continuum intensity $I$ is low, $CF > 1$ and the missing field strength is recovered; in bright features $CF < 1$, preventing overcorrection. Typical $CF$ values range from 0.5 to 2.5, demonstrating the pixel-wise adaptability of the LIC method.

The improvement is particularly evident in the umbral cores. For AR24051, the uncorrected field ranges from $-303$ to $+381$\,G, while after correction it expands to $-471$ to $+587$\,G. For the stronger AR24153, the enhancement is more substantial: $B_{L0}$ ranges from $-486$ to $+476$\,G, whereas $B_L$ extends from $-735$ to $+812$\,G---a gain of over 70\% in the positive polarity peak. These increases reflect the recovery of magnetic flux in the darkest umbrae, where the linear calibration most severely underestimates the true field. The originally weak and diffuse magnetic signals in $B_{L0}$ are significantly enhanced in $B_L$, bringing the umbral fields into closer visual agreement with the expected morphology of sunspot magnetic structures. This behavior is consistently observed across the entire sample of 26 observations: the positive polarity peak enhancement ranges from $\sim$90 to 668\,G, with a median enhancement of approximately 310\,G, and the negative polarity peak shows similar improvement, with enhancements ranging from $\sim$90 to 488\,G and a median enhancement of approximately 260\,G. These results indicate that the LIC correction effectively compensates for the underestimation across a wide range of field strengths and active-region configurations.

\begin{figure}
\centering
\includegraphics[width=\linewidth]{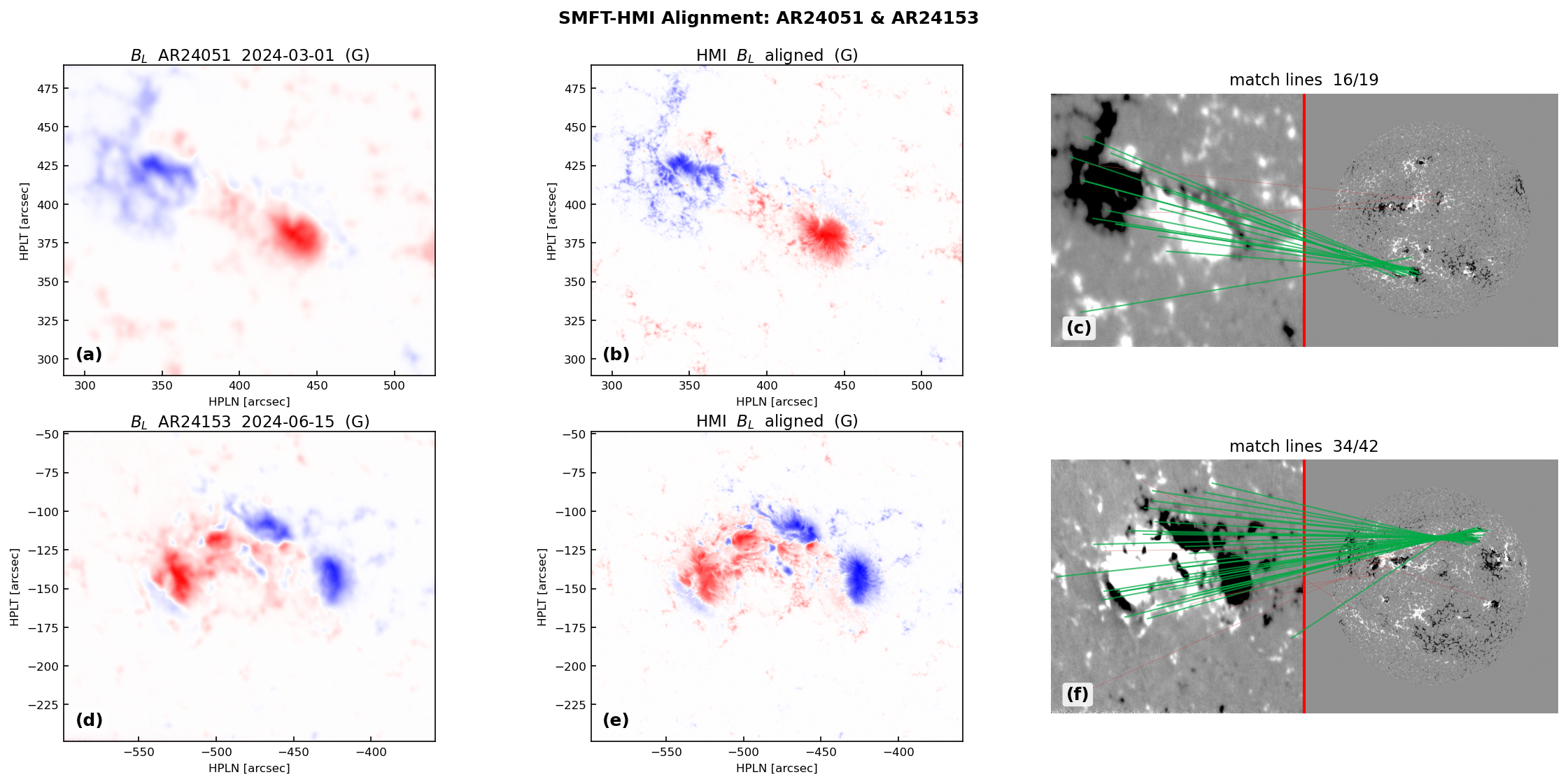}
\caption{SIFT+RANSAC registration for AR24051 (top) and AR24153 (bottom).
Left: SMFT $B_L$ magnetograms. Middle: registered HMI subframes. Right:
SIFT keypoint matches, with green lines denoting RANSAC inliers.}
\label{fig:align}
\end{figure}

The pixel-wise comparison between SMFT and HMI requires precise spatial alignment of the two datasets. To achieve this, we employ SIFT (Scale-Invariant Feature Transform) keypoint detection and matching, followed by RANSAC (Random Sample Consensus) estimation of an affine homography that maps the HMI image onto the SMFT coordinate system. Figure~\ref{fig:align} illustrates the registration procedure and results for the same two active regions shown in Figure~\ref{fig:bl_cf}: AR24051 (top) and AR24153 (bottom). The figure consists of three panels for each active region. The left panels display the SMFT $B_L$ magnetograms of the active region. The middle panels show the corresponding HMI subframes cropped from the full-disk magnetograms after registration, which are used for the subsequent pixel-wise comparison. The right panels present the SIFT matching process: the left subpanel within this panel is the SMFT $B_L$ image, and the right subpanel is the HMI full-disk magnetogram, with green lines connecting the matched SIFT keypoints that are identified as geometrically consistent inliers by the RANSAC algorithm. The dense distribution of these valid matching points across the field of view demonstrates that the SIFT detector successfully identifies abundant and reliable features even in the relatively low-contrast SMFT magnetograms. The affine homography derived from these inlier matches achieves accurate registration, as evidenced by the visual alignment of sunspot umbrae and surrounding magnetic structures between the two instruments in the left and middle panels. This high-quality alignment is quantitatively supported by RANSAC inlier fractions ranging from 40\% to 88\% across all successfully registered observations, indicating that a large majority of the SIFT keypoint matches are geometrically consistent and suitable for homography estimation. The robust performance of the SIFT+RANSAC pipeline across active regions with varying magnetic complexity and contrast confirms its suitability for routine application in SMFT--HMI cross-calibration studies.

\begin{figure}
\centering
\includegraphics[width=\linewidth]{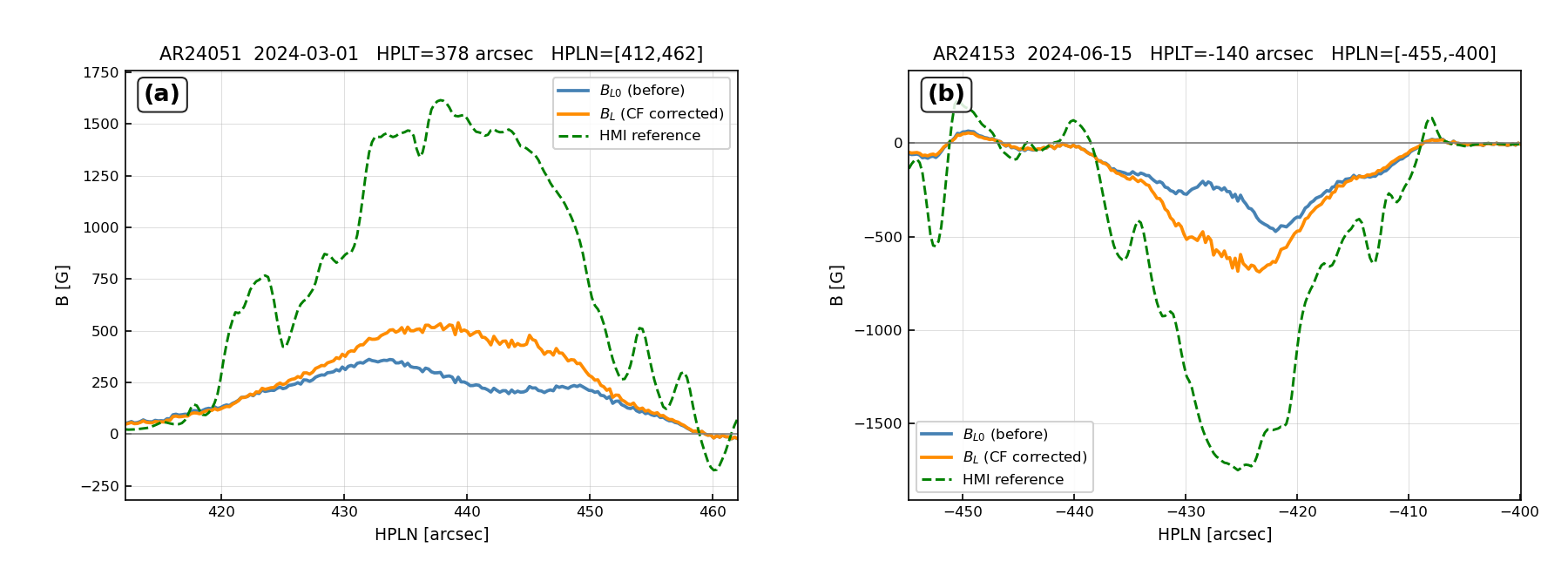}
\caption{Horizontal cut profiles of $B_{L0}$ (blue), $B_L$ (orange), and the
HMI reference (green dashed) across the sunspot centers of AR24051 (left)
and AR24153 (right). Cut positions are marked in Figure~\ref{fig:bl_cf}.}
\label{fig:linecut}
\end{figure}

To further illustrate the effect of the LIC correction at the umbral scale, we extract horizontal cuts across the sunspot centers of AR24051 and AR24153. The locations of these cuts are indicated by the lime-colored line segments overlaid on the left panels ($B_{L0}$) of Figure~\ref{fig:bl_cf} (top: AR24051; bottom: AR24153). Figure~\ref{fig:linecut} shows the corresponding profiles of $B_{L0}$, $B_L$, and the HMI reference along these cuts, with the cut coordinates (HPLT and HPLN ranges) annotated in each subplot. In the uncorrected field $B_{L0}$ (blue), the umbral cores exhibit a pronounced dip or flattening, where the magnetic signal is anomalously weak compared to the surrounding penumbral regions. This dip reflects the magnetic underestimation inherent to the single-wavelength linear calibration in the darkest umbrae. After LIC correction (orange), the profiles are substantially elevated in the umbral cores, transforming the dip into a well-defined peak that closely follows the morphology of the HMI reference (green dashed). This recovery of the umbral peak is consistent with the expected magnetic field configuration of sunspots, where the strongest fields are located in the darkest central umbrae. The close agreement between the corrected SMFT profiles and HMI demonstrates that the LIC correction effectively restores the physically expected umbral field enhancement, removing the artificial depression introduced by the linear calibration.

\begin{figure}
\centering
\includegraphics[width=\linewidth]{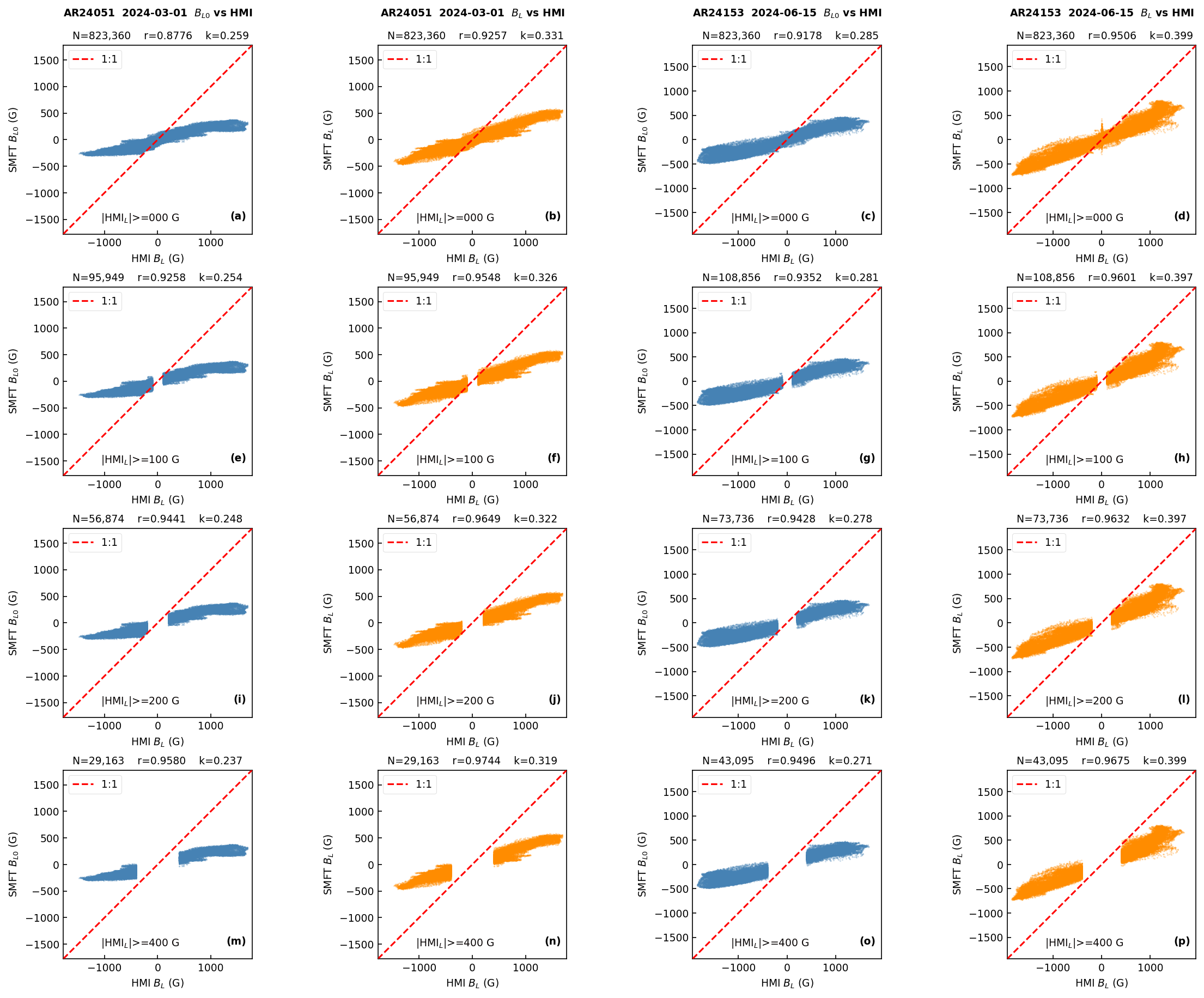}
\caption{Pixel-by-pixel scatter diagrams of SMFT versus HMI longitudinal field
for AR24051 (left two columns) and AR24153 (right two columns) at four
threshold levels (top to bottom): All, $>100$, $>200$, and $>400$\,G.
Blue: uncorrected $B_{L0}$; orange: LIC-corrected $B_L$; dashed red line: 1:1.}
\label{fig:scatter}
\end{figure}

To quantify the improvement, we perform pixel-by-pixel scatter analysis. Figure~\ref{fig:scatter} shows scatter diagrams for AR24051 (left two columns) and AR24153 (right two columns) at four threshold levels (All, $>100$, $>200$, $>400$\,G). The uncorrected field $B_{L0}$ (blue) systematically deviates from the 1:1 line (dashed red), particularly in strong-field regions, reflecting the well-known magnetic underestimation inherent to the single-wavelength linear calibration. After LIC correction (orange), the distribution shifts systematically toward the 1:1 line, indicating that the correction successfully recovers a substantial portion of the missing field strength. Although the SMFT measurements remain systematically lower than HMI even after correction---a residual offset that suggests the SMFT calibration constant may need further refinement---the LIC-corrected points are consistently closer to the 1:1 line than their uncorrected counterparts, particularly in the strong-field regimes. Quantitatively, the regression slope $k$ increases from $\approx0.24$--0.26 for $B_{L0}$ to $\approx0.32$--0.40 for $B_L$, and the correlation coefficient $r$ increases in every panel. The improvement is most pronounced at higher thresholds ($>200$ and $>400$\,G), where the underestimation is most severe and the correction therefore has the greatest impact. The consistency of the improvement across all four threshold levels further demonstrates the robustness of the LIC method.

\begin{figure}
\centering
\includegraphics[width=\linewidth]{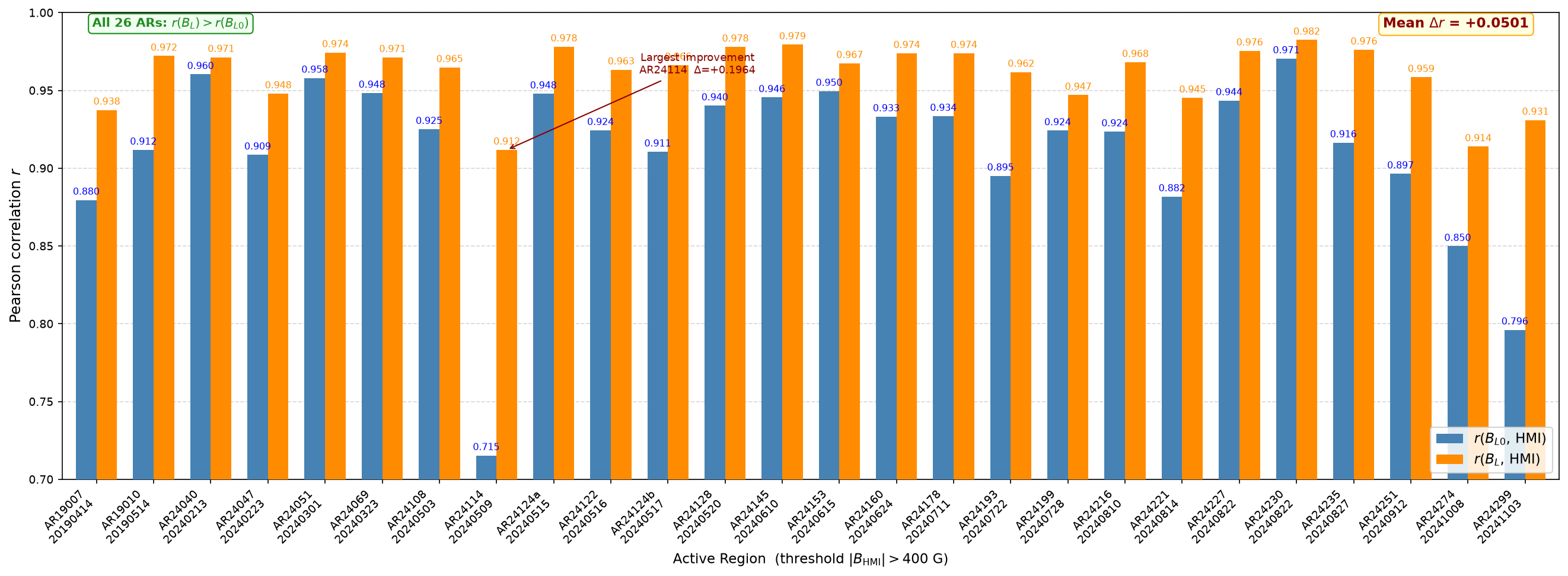}
\caption{Pearson correlation coefficient $r$ between SMFT and HMI at
$|B_{\rm HMI}|>400$\,G for all 26 active-region observations. Blue bars:
LIC-corrected $B_L$; orange bars: uncorrected $B_{L0}$.}
\label{fig:corr_all}
\end{figure}

Figure~\ref{fig:corr_all} presents the Pearson correlation coefficient $r$ between SMFT and HMI for all 26 active-region observations at the $|B_{\rm HMI}|>400$\,G threshold, providing a global assessment of the LIC correction performance across our entire sample. For every one of the 26 cases, the LIC-corrected field $B_L$ (blue bars) yields a higher correlation coefficient than the uncorrected $B_{L0}$ (orange bars). This consistent improvement is particularly noteworthy given the diversity of the sample, which includes active regions spanning a wide range of magnetic complexity and evolutionary stages. The improvement varies considerably across the sample, with a mean of $\Delta r=+0.041$, a maximum of $\Delta r=+0.197$ (AR24114, from 0.715 to 0.912), and a minimum of $\Delta r=+0.012$ (AR24230, from 0.971 to 0.982). The maximum improvement of nearly 0.2 in correlation coefficient represents a substantial gain, transforming a moderately correlated field into one that is strongly correlated with the HMI reference. In contrast, the minimum improvement is observed in a case where the uncorrected field already exhibited an exceptionally high correlation ($r=0.971$), leaving little room for further enhancement. The fact that the LIC correction improves the correlation in all 26 cases considered here---regardless of the initial correlation level or the diversity of active region characteristics---is a strong indication that the method is broadly applicable across a wide range of solar conditions,although this sample does not exclude the possibility of failure in special cases not represented here.

\section{Discussion}
\label{sec:discuss}

\subsection{Effectiveness of the LIC correction}

The improvement across all 26 diverse active regions indicates that the LIC correction is both generally applicable and physically sound. The method relies solely on the local continuum intensity $I$, which is directly measured by SMFT, and requires no external calibrator, no reference magnetogram, and no explicit underestimation threshold. It is therefore well suited for automated pipeline processing of the HSOS multi-decadal data archive (1987--2024).

The mean improvement of $\Delta r=+0.041$ confirms that the LIC correction provides a statistically significant enhancement in correlation with HMI, consistent with the level of improvement expected from practical calibration efforts \citep{2021RAA....21...67X,2021SoPh..296..165P}, although the specific metrics differ between studies. The improvement varies across the sample, with AR24114 (NOAA 13667) showing the largest gain of $\Delta r=+0.197$, where the correlation increases from 0.715 to 0.912. This indicates that the correction can be particularly effective in cases where the initial correlation is moderate and the magnetic underestimation is more pronounced, which is exactly where the correction has the greatest practical value for quantitative applications such as helicity estimation, flux measurement, and helioseismology. Conversely, active regions with already high initial correlation, such as AR24230 ($r_{B_{L0}}=0.971$), exhibit more modest improvements, as there is limited room for further enhancement.

The sample spans a diverse range of active regions, from weak simple bipolar configurations to complex multi-spot groups, with sunspot umbrae of varying sizes and magnetic field strengths, and covers both solar minimum and solar maximum periods. Despite this diversity, the correction remains consistently effective, demonstrating that the LIC method is robust and does not depend on specific active-region characteristics. This consistency is particularly encouraging for future applications, as it suggests that the LIC can be reliably applied to the full HSOS data archive without the need for case-by-case tuning or external calibration. Because the method is fully data-driven and requires only the observed continuum intensity $I$---which is already available in the standard SMFT data products---it can be easily integrated into existing data reduction pipelines with minimal additional computational cost. Furthermore, since the correction is applied pixel-wise and does not depend on any external reference, it can be consistently applied to historical observations for which contemporaneous HMI or other reference data may not be available. This feature is particularly valuable for long-term studies that aim to utilize the full multi-decadal HSOS archive to investigate solar cycle variations of photospheric magnetic fields and their associated flare activity. The overall improvement in correlation, together with the improvement across all 26 active regions, provides strong quantitative evidence that the LIC correction is a valuable and reliable preprocessing step for SMFT longitudinal magnetograms.

\subsection{Limitations of the single-wavelength approach}

\begin{figure}
\centering
\includegraphics[width=\linewidth]{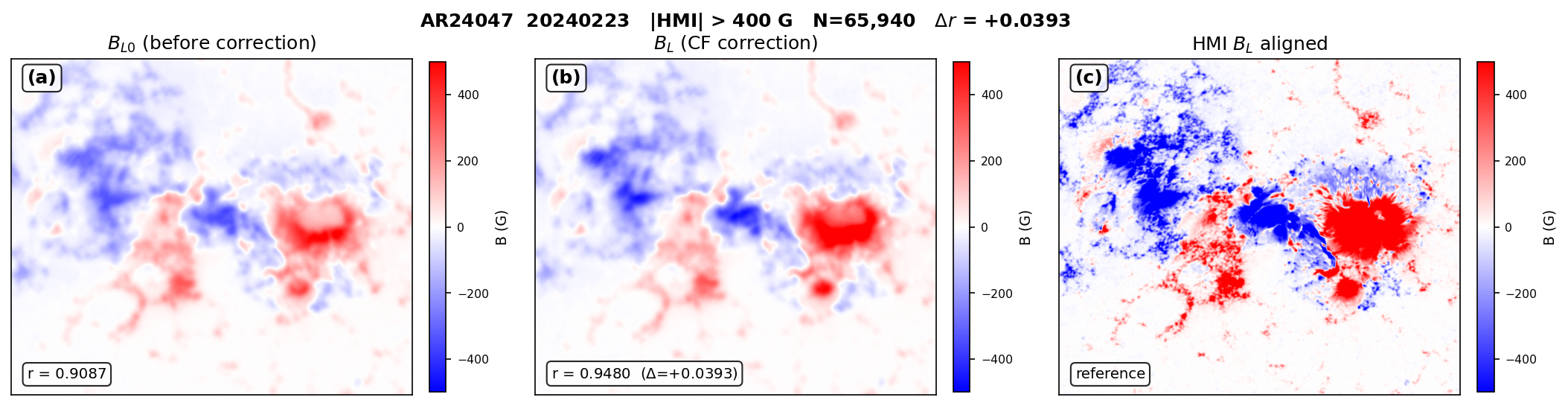}
\caption{Three-panel comparison for AR24047 (2024 February 23): uncorrected
$B_{L0}$ (left), LIC-corrected $B_L$ (middle), and HMI reference (right).}
\label{fig:sat24047}
\end{figure}

\begin{figure}
\centering
\includegraphics[width=\linewidth]{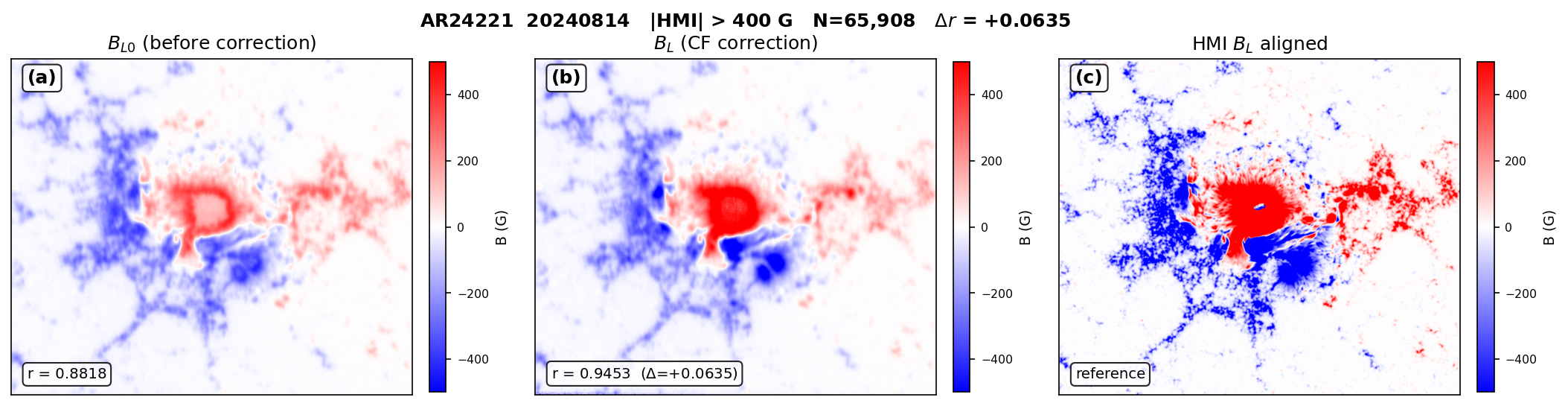}
\caption{Same as Figure~\ref{fig:sat24047}, but for AR24221 (2024 August 14).}
\label{fig:sat24221}
\end{figure}

Despite the overall improvement demonstrated in the previous sections, the LIC correction cannot fully recover the magnetic field in the most severely underestimated pixels due to the inherent limitation of the single-wavelength linear calibration. Figures~\ref{fig:sat24047} and~\ref{fig:sat24221} illustrate this residual limitation through three-panel comparisons for two representative active regions, AR24047 and AR24221, both of which contain strong sunspot umbrae where the single-wavelength underestimation is most pronounced. In each figure, the left panel shows the uncorrected SMFT $B_{L0}$ magnetogram, the middle panel displays the LIC-corrected $B_L$ field, and the right panel presents the corresponding HMI reference magnetogram.

For both active regions, the correction substantially enhances the magnetic signals in the sunspot umbrae, recovering a significant portion of the field strength that was underestimated in the original $B_{L0}$. The umbral regions, which appear notably weaker and more diffuse in the uncorrected magnetograms, are visibly strengthened in the corrected versions. However, even after correction, the central umbral cores in the SMFT magnetograms remain weaker than the surrounding penumbral regions, while the HMI reference shows the opposite pattern---stronger fields in the umbrae and weaker fields in the penumbrae. This visual discrepancy suggests that the intensity-based correction, while effective in recovering a substantial portion of the missing field, cannot completely eliminate the underestimation in the darkest umbral cores where the Stokes $V$ signal is most severely affected. The residual discrepancies are particularly evident in these two active regions clearly demonstrating the same fundamental limitation.

Upon closer inspection of the HMI panels, some anomalies can also be seen in the sunspot centers. In both figures, the HMI magnetograms exhibit localized weakening or irregular structures in the strongest umbral cores, precisely where the continuum intensity is lowest. Consequently, the residual differences between SMFT and HMI in the umbral centers cannot be attributed solely to the SMFT linear calibration; a portion of the discrepancy may reflect inherent uncertainties in the HMI inversion as well. Nevertheless, the consistency of the residual underestimation pattern across multiple active regions strongly indicates that the primary limitation is the single-wavelength calibration of SMFT.

The fundamental limitation is inherent to the single-wavelength formulation. Once the Stokes $V$ signal becomes fully nonlinear with respect to the magnetic field in the strongest umbrae, the local continuum intensity alone provides insufficient information to uniquely recover the true field strength. Multi-wavelength approaches that fit the full line profile can break this degeneracy but require the multi-point scanning mode of SMFT, which is not available for the routine single-wavelength observations used throughout this study.

\subsection{Implications of the LIC Correction}
\label{sec:implications}

Before the correlation analysis, the HMI magnetograms were degraded to
the SMFT spatial resolution, so that $B_{L0}$, $B_L$, and the HMI
reference are compared at a matched resolution. The LIC correction acts
mainly on the dark sunspot umbrae, where the field underestimation is
most severe, while its effect on small-scale structures and on the
penumbra is relatively minor.

\begin{figure}
\centering
\includegraphics[width=\linewidth]{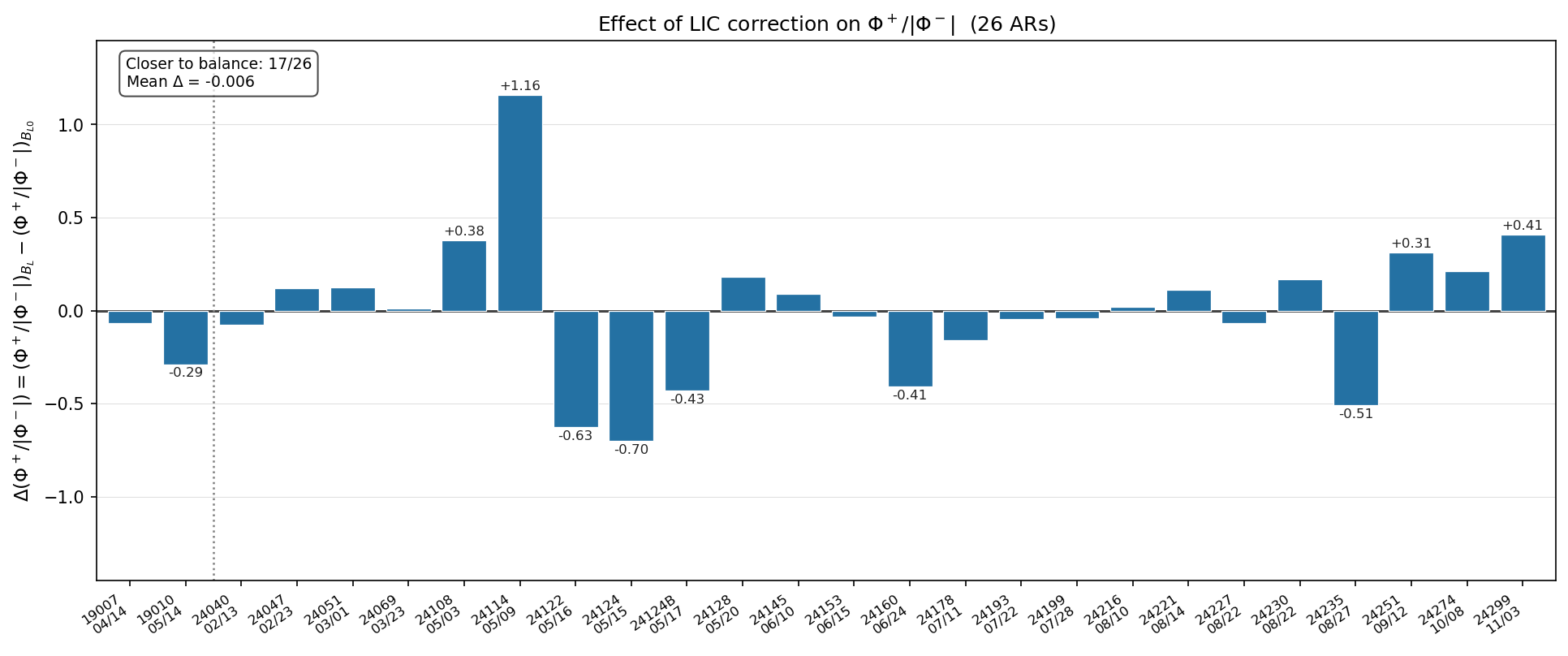}
\caption{Change of the magnetic flux balance ratio $\Phi^+/|\Phi^-|$ after
the LIC correction for the 26 active regions.}
\label{flux_balance_ratio}
\end{figure}

One example of the practical impact of the corrected data concerns
the flux balance between the leading and following polarities. In our
sample, the enhanced flux comes mainly from the leading-spot umbrae,
whereas the following spots rarely show a pronounced field
underestimation; the corrected magnetograms therefore tend to increase
the apparent leading--following flux imbalance. Part of this change may
reflect real physical differences between the two polarities, but part
may also be introduced by the correction itself. A systematic check of
the flux imbalance before and after the correction, and of its
dependence on the leading--following configuration, is therefore needed
before the corrected data are used in studies that are directly sensitive
to the absolute flux balance. Figure~\ref{flux_balance_ratio} compares 
the flux balance ratio $\Phi^+/|\Phi^-|$ before and after the LIC
correction for all 26 active regions.  A negative $\Delta$ indicates that the
correction reduces the positive flux relative to the negative 
(i.e., enhances the leading-polarity dominance), whereas a positive
value indicates the opposite.  The changes are mostly modest: for the 
24 active regions in 2024 the mean $|\Delta| = 0.21$ and the median $|\Delta| = 0.12$.  
Eleven of the 26 active regions show $|\Delta| < 0.05$, indicating that the 
leading and following polarities are already close to balance and the 
correction has little effect on the flux ratio.  Two ARs stand out with 
$|\Delta| > 0.45$ (ar24114: $\Delta = +1.16$; ar24124: $\Delta = -0.70$) 
because they are strongly unbalanced even before correction 
($\Phi^+/|\Phi^-| \approx 4.1$ and $2.4$, respectively) and the 
correction primarily removes spurious positive flux from the leading 
umbrae. The two 2019 active regions also display comparatively large changes 
(ar19007: $\Delta = -0.07$; ar19010: $\Delta = -0.29$).

\section{Conclusion}
\label{sec:conclude}

Using SIFT+RANSAC image registration we established pixel-level co-registration
between SMFT active-region cutouts and co-temporal SDO/HMI full-disk
magnetograms for 26 active regions observed in 2019 and 2024.  We validated the
Local Intensity-based Correction (LIC) implemented in the HSOS standard pipeline and found:

\begin{enumerate}

\item The LIC correction improves the Pearson correlation coefficient $r$ with
HMI for all 26 active regions considered here. The mean
improvement is $\Delta r=+0.041$, with a maximum of $\Delta r=+0.197$
(AR24114) and a minimum of $\Delta r=+0.012$ (AR24230). The correction
is effective across all types of active regions in our sample, requiring
only the observed continuum intensity $I$ itself---already available in
the standard SMFT data products---without any external calibration or
additional observational constraints.

\item Residual errors persist in the darkest umbral cores, confirming an
inherent limitation of any single-wavelength linear calibration approach.
The HMI reference itself may also exhibit anomalies in the strongest umbrae due to limitations in the inversion model and instrument resolution.

\item We recommend that all future quantitative applications of SMFT longitudinal
magnetograms apply the LIC correction as a standard preprocessing step.
An automated pipeline for applying the LIC across the full HSOS archive
is under development. In addition, we plan to extend this method to
data from other similar filter-type single-wavelength magnetographs that
suffer from comparable effects.

\end{enumerate}

\begin{acknowledgements}
This research was supported by the National Key R\&D Program of China (No. 2022YFF0503001, 2021YFA1600500), National Natural Science Foundation of China (grant Nos. 12473052, 12273059 and 12473053),  China's Space Origins Exploration Program (GJ11020400), the Strategic Priority Research Program on Space Science of the Chinese Academy of Sciences (Grant No. XDA15320000), and the specialized research fund for State Key Laboratory of Solar Activity and Space Weather.
\end{acknowledgements}

\bibliography{mybib}
\bibliographystyle{aasjournalv7}

\end{document}